\documentclass[prb,twocolumn,superscriptaddress,showpacs]{revtex4}
\usepackage{graphicx}
\usepackage{latexsym}
\newcommand{\fig}[2]{\includegraphics[width=#1]{#2}}

\begin{document}
\renewcommand{\ni}{{\noindent}}
\newcommand{\dprime}{{\prime\prime}}
\newcommand{\be}{\begin{equation}}
\newcommand{\ee}{\end{equation}}
\newcommand{\bea}{\begin{eqnarray}}
\newcommand{\eea}{\end{eqnarray}}
\newcommand{\nn}{\nonumber}
\newcommand{\bk}{{\bf k}}
\newcommand{\bK}{{\bf K}}
\newcommand{\bQ}{{\bf Q}}
\newcommand{\bN}{{\bf \nabla}}
\newcommand{\bA}{{\bf A}}
\newcommand{\bE}{{\bf E}}
\newcommand{\bj}{{\bf j}}
\newcommand{\bJ}{{\bf J}}
\newcommand{\bs}{{\bf v}_s}
\newcommand{\bn}{{\bf v}_n}
\newcommand{\bv}{{\bf v}}
\newcommand{\la}{\langle}
\newcommand{\ra}{\rangle}
\newcommand{\dg}{\dagger}
\newcommand{\br}{{\bf{r}}}
\newcommand{\brp}{{\bf{r}^\prime}}
\newcommand{\bq}{{\bf{q}}}
\newcommand{\hx}{\hat{\bf x}}
\newcommand{\hy}{\hat{\bf y}}
\newcommand{\bS}{{\vec S}}
\newcommand{\bsigma}{{\vec{\sigma}}}
\newcommand{\cU}{{\cal U}}
\newcommand{\cD}{{\cal D}}
\newcommand{\bR}{{\bf R}}
\newcommand{\pll}{\parallel}
\newcommand{\sumr}{\sum_{\vr}}
\newcommand{\cP}{{\cal P}}
\newcommand{\cQ}{{\cal Q}}
\newcommand{\cS}{{\cal S}}
\newcommand{\upa}{\uparrow}
\newcommand{\dna}{\downarrow}
\title{Effect of local charge fluctuations on spin
physics in the N\'eel state of La$_2$CuO$_4$}
\author{Luca Capriotti}
\affiliation{Credit Suisse First Boston (Europe) Ltd.,
One Cabot Square, London E14 4QJ, United Kingdom}
\author{Andreas L\"auchli}
\affiliation{Institut Romand de Recherche Num\'erique en Physique des 
Mat\'eriaux (IRRMA),
  PPH-Ecublens, CH-1015 Lausanne
}
\author{Arun Paramekanti}
\affiliation{Department of Physics, University of California, Berkeley
94720-7300, U.S.A.}
\begin{abstract}
\vspace{0.1cm}
We explore the effect of local charge fluctuations on the spin response
of a Mott insulator by deriving an effective spin model, and studying
it using Schwinger boson mean field theory. Applying this to 
La$_2$CuO$_4$, we show that an accurate fit to the magnon dispersion relation,
measured by Coldea {\em et al.} [Phys. Rev. Lett. {\bf 86}, 5377 (2001)]
is obtained with Hubbard model parameters $U \approx 2.34 eV$,
and $t \approx 360 meV$. These parameters lead to estimates of the
staggered magnetization ($m_s \approx 0.25$), spin wave
velocity ($c\approx 800 meV$-\AA), and spin stiffness
($\rho_s \approx 24 meV$). In particular the staggered moment as
well as the effective local moment are renormalized to smaller values
compared to the Heisenberg model due to local charge fluctuations in 
the Hubbard model.  The dynamical structure factor shows considerable 
weight in the continuum along the zone boundary as well as secondary peaks 
that may be observed in high resolution neutron scattering experiments.
\end{abstract}
\maketitle
\section{Introduction}
Multispin interactions are important in a variety of magnetic
systems.\cite{intromat} In a magnetically ordered insulator, such
interactions can reveal themselves through their effect on the
spin dynamics measured in neutron scattering experiments. One
system which appears to provide an example of such physics is
La$_2$CuO$_4$, which is a Mott insulating antiferromagnet. Indeed,
high resolution data on the magnon dispersion in this system
indicate a dip in the magnon energy near $\bQ=(\pi/2,\pi/2)$ while
traversing the magnetic Brillouin zone boundary along $(\pi,0)\!\!
\to \!\!(0,\pi)$. Such a dip is not expected for the magnon
dispersion in a nearest-neighbor Heisenberg antiferromagnet, but
it has been suggested \cite{coldea} that this could arise from
four-spin interactions generated by local charge fluctuations in
the insulating phase. Motivated by this, we revisit the zero
temperature N\'eel ordered state of
La$_2$CuO$_4$, and examine how the spin stiffness, the ordered moment, 
the effective local moment at a site, and the spin dynamics are
influenced by local charge fluctuations in this Mott insulator.

La$_2$CuO$_4$ is a layered antiferromagnet. While
the exchange coupling between the two-dimensional layers is
crucial for the existence of a finite-temperature N\'eel
transition in this system, this interplane coupling is
nevertheless tiny in magnitude, $\sim 10^{-5}$ of the in-plane
exchange coupling, and much smaller than the resolution of the
neutron scattering experiments. It is therefore expected that the
low temperature spin dynamics in the N\'eel ordered state in this
material is adequately captured by the Hubbard model at zero
temperature on a two dimensional square lattice, (with standard
notations)
\begin{equation}
H = - t \sum_{\la i j\ra, \sigma} \left( c^\dagger_{i\sigma}
c^{\vphantom\dagger}_{j\sigma} + h.c.\right) + U \sum_i n_{i\upa} 
n_{i\dna}
\label{hubbard}
\end{equation}
in its insulating phase at strong coupling $U \gg t$. In this
regime, it is well known\cite{takahashi,macdonald} that the low
energy physics can be encapsulated in an effective spin
Hamiltonian which incorporates two-spin and four-spin interactions
(given below in Eq.~\ref{hspin}).

A detailed analysis of a spin model with such
four-spin interactions was carried out by Katanin and Kampf \cite{kampf}
using modified spin wave theory. This allowed them to conclude
that a fit to the magnon dispersion indeed requires sizeable
four-spin interactions, as originally suggested by Coldea
{\it et al} \cite{coldea} based on linear spin wave theory.
A different approach
to this problem was adopted by Singh and Goswami \cite{goswami}
and Peres and Ara\'ujo \cite{peres} who used
the random-phase approximation (RPA) for the Hubbard model
with a spin density wave ground state.
This is presumably a better approximation at weak coupling but
nevertheless also reproduces the dip in the magnon
dispersion near $(\pi/2,\pi/2)$ at intermediate values of $U/t$.
 
In this paper, we use Schwinger boson mean field theory to study
the magnon dispersion and scattering continuum for the effective
spin model in Eq.~(\ref{hspin}). There are several reasons to try
this route. First, in contrast to the RPA studies, this approach
of mapping to an effective spin model is explicitly geared towards 
addressing the strong coupling limit $U/t \gg 1$.
It is therefore interesting to compare the two approximations
at the intermediate coupling values of experimental interest,
especially taking into account modifications to the spin operator
which result when going from the Hubbard electronic description
to an effective spin model.
Indeed, it is satisfying that some of the results obtained here are 
consistent
with the spin-density wave approach for intermediate $U/t$.
Second, Schwinger boson mean field theory is known to work
well for the square lattice Heisenberg model and it is
worth asking if it continues to be reliable once sizeable
four-spin interactions are included, and how such interactions
affect the ground state properties and the magnon dispersion.
Finally, we consider the scattering continuum beyond the magnon peak 
in La$_2$CuO$_4$
which has not been addressed sufficiently in these earlier papers.
(While this paper was being refereed, a closely related
work \cite{lorenzana} which discusses the sum-rule and other
issues with neutron scattering experiments in La$_2$CuO$_4$
was submitted for publication. We have therefore added
a brief discussion of sum-rules within our approach, and refer 
the reader to the above paper for a detailed analysis of
sum rules, the current status of neutron scattering experiments
and alternative theoretical approaches to spin dynamics in this 
system.)

A summary of our main new results is given below.

(i) Comparing the mean field
theory results for the ground state energy and static spin
structure factor with exact diagonalization studies on small
clusters, we show that the mean field theory provides a good
description of the ground state of the spin model (\ref{hspin}),
in the experimentally relevant regime. 

(ii) We show that the
experimental magnon dispersion is well reproduced by our approach
for Hubbard model parameters $U \approx 2.34 eV$, and $t \approx
360 meV$. These parameter values are in reasonable agreement with
the experimental estimates \cite{optics} of $U$, and electronic
structure calculations \cite{estruct} for $t$. 
Our estimated ratio $U/t
\approx 6.5$ is in rough agreement with the value inferred from
RPA calculations for the Hubbard model \cite{goswami,peres}
as well as the quantum Monte Carlo (QMC) calculations of Sengupta {\em et
al.}, \cite{sengupta} though we do not resort to the single mode
approximation used by the latter authors. It is however slightly
smaller than the ratio $U/t \approx 8.8$ obtained from a recent
series expansion study of the Hubbard model \cite{zheng05}.

(iii) Our best fit
parameters for $t,U$ lead to values for the staggered magnetization $m_s
\approx 0.25$, spin wave velocity $c\approx 800~meV$-$\AA$, and
spin stiffness $\rho_s \approx 24 meV$. The spin wave velocity and
spin stiffness are in good agreement with experimental findings.
\cite{neutrons} The staggered
moment is substantially reduced from that of the Heisenberg model due 
to local charge fluctuations, as was earlier pointed out by Delannoy 
{\it et al}, \cite{tremblay} and it may be worth revisiting this in
experiments.

(iv) Turning to the spin excitation spectrum beyond the magnon,
we find that there is a broad continuum of excitations at 
energy scales relevant to neutron scattering experiments. In
addition, the spin-flip operator in the Hubbard model induces
transitions into the upper Hubbard band which would show up at
energies of ${\cal O}(U)$. We however restrict attention in this
paper to energies well below the Mott gap, where a spin model 
description is appropriate. In this regime, we find
that, along the zone boundary, the spectral weight in the
continuum is about $40\%$ of the total spectral weight. This is
similar to the large continuum spectral weight inferred
from QMC calculations on the Heisenberg model,\cite{sandvik}
but we cannot make any quantitative comparisons at this stage. 
Turning to the
spectral lineshape at energy scales relevant to neutron
scattering experiments, we find that the dynamical structure
factor exhibits secondary peaks within the mean field theory. 

(v) The local charge
fluctuations in the Hubbard model also lead to a smaller effective 
local moment. To
obtain this effective moment, we consider integrating the
dynamical structure factor over energies well below the Mott gap,
as relevant for neutron scattering experiments, and then
integrating this ``low-energy'' equal-time structure factor 
$\tilde{\cS}_{\rm low}(\bq)$ over $\bq$. This leads to the sum-rule 
$\sum_\bq \tilde{\cS}_{\rm low}(\bq) = S_{\rm eff} (S_{\rm eff}+1)$, 
with $S_{\rm eff} < S = 1/2$. Our calculation gives $S_{\rm eff} 
\simeq 0.39$.

It is possible that all these features could be explored in 
neutron scattering experiments with higher resolution and intensity 
in the near future.

\section{Effective spin model}
The effective spin model describing the strong coupling regime
of the Hubbard model in (\ref{hubbard}) takes the form
\bea
\!\!H_{\rm spin} \!\!&=&\!\!\! \frac{1}{2} \sum_{i,j} J(\br_i-\br_j)
\bS_i \cdot \bS_j
+ \sum_{\Box} J_{\Box} \left[(\bS_1\cdot\bS_2) (\bS_3\cdot\bS_4) \right. 
\nn\\
&+& \left.
(\bS_1\cdot\bS_4) (\bS_2\cdot\bS_3) - (\bS_1\cdot\bS_3) (\bS_2\cdot\bS_4)
\right].
\label{hspin}
\eea
Here, $\bS_i$ are spin-half operators, and $J_\Box$ refers to an 
elementary plaquette on
the square lattice with $1-4$ labeling sites on its corners.
The explicit expressions for the exchange
couplings in terms of the Hubbard model parameters are\cite{macdonald},
\bea
J_1 &=& J_{\hat{x}}=J_{\hat{y}}=4 \frac{t^2}{U} - 24 \frac{t^4}{U^3} \\
J_2 &=& J_{\hat{x}+\hat{y}}=J_{\hat{x}-\hat{y}}= 4 \frac{t^4}{U^3} \\
J_3 &=& J_{2\hat{x}}=J_{2\hat{y}}= 4 \frac{t^4}{U^3} \\
J_\Box&=&80 \frac{t^4}{U^3}
\label{Jexpr}
\eea
With hopping matrix elements between further neighbor sites in the
Hubbard model, we expect additional couplings
in the effective spin Hamiltonian. However,
electronic structure calculations \cite{estruct} suggest that
these matrix elements are small in magnitude
for La$_2$CuO$_4$, so we will not consider them here.
\section{Schwinger boson mean field theory}
To study the ground state and excitations of $H_{\rm spin}$, we
follow Ref.~\onlinecite{sbreview} and
represent spins using two species of bosons $b_{1,2}$ as
\be
\bS_i = b^\dagger_{i\alpha} \bsigma_{\alpha\beta}
b^{\vphantom\dagger}_{i\beta}
\label{sbrepresentation}
\ee
with the constraint
\be
b^{\dagger}_{i\alpha} b^{\vphantom\dagger}_{i\alpha} = 2S
\label{constraint}
\ee
on the boson number at each site. The $\bsigma$ are Pauli matrices,
the spin $S=1/2$ for our system and, unless specified, the summation over
repeated Greek indices
is implied here and below. The Hamiltonian can be reexpressed in terms
of the bosons by using the identity
\be
\bS_i \cdot \bS_j = :B^\dagger_{ij} B^{\vphantom\dagger}_{ij}: -
A^\dagger_{ij} A^{\vphantom\dagger}_{ij}~,
\ee
where $: :$ is the standard normal ordering and the bond operators are
\bea
B_{ij}&=& \frac{1}{2} b^\dagger_{i\alpha} b^{\vphantom\dagger}_{j\alpha} 
\\
A_{ij}&=& \frac{1}{2} (b^{\vphantom\dagger}_{i,1} 
b^{\vphantom\dagger}_{j,2} -
b^{\vphantom\dagger}_{i,2} b^{\vphantom\dagger}_{j,1}).
\eea
In the exact Schwinger boson representation, the Hamiltonian obtained 
using
the above identity
consists of 4-boson and 8-boson operators, and one has in addition to deal 
with the
boson number constraint (\ref{constraint}). In order to make progress we 
can
resort to a mean field approach.
Applied to the Hamiltonian $H_{\rm spin}$ in Eq.~(\ref{hspin}),
this consists of three approximations.
(i) First,
we reduce the 4-spin interactions to effective 2-spin terms by
setting
\bea
(\bS_1\cdot\bS_2)
(\bS_3\cdot\bS_4) &\to& \la (\bS_1\cdot\bS_2) \ra (\bS_3\cdot\bS_4) \\
&+& (\bS_1\cdot\bS_2) \la (\bS_3\cdot\bS_4) \ra \\
&-& \la (\bS_1\cdot\bS_2) \ra \la (\bS_3\cdot\bS_4) \ra.
\eea
This renormalizes the two-spin exchange couplings
$J(\br_i -\br_j) \to J^{\rm eff} (\br_i-\br_j)$ in $H_{\rm spin}$.
It is straightforward to show that $J^{\rm eff}_1 = J_1 +
2 J_{\Box} \la \bS_i\cdot \bS_j\ra$ where $ij$ are nearest
neighbor spins, $J^{\rm eff}_2 = J_2 - J_{\Box}
\la \bS_i \cdot \bS_m\ra$ where $im$ are next nearest neighbors,
and $J^{\rm eff}_3 = J_3 $. At this stage, we
obtain a Heisenberg spin Hamiltonian $H^{\rm mf}_{\rm spin}$
with only two-spin interactions.
(ii) Next, we do a mean field decoupling of the 4-boson terms arising
from the effective two-spin interactions,
setting $:B^\dagger_{ij} B^{\vphantom\dagger}_{ij}: \to
\la B^\dagger_{ij} \ra B^{\vphantom\dagger}_{ij}
+ B^\dagger_{ij} \la B^{\vphantom\dagger}_{ij}\ra
- \la B^\dagger_{ij} \ra \la B^{\vphantom\dagger}_{ij}\ra$ and similarly
for $A^\dagger_{ij} A^{\vphantom\dagger}_{ij}$. This reduces the
spin Hamiltonian $H^{\rm mf}_{\rm spin}$ to a quadratic boson
Hamiltonian, which up to overall constant terms is given by
\begin{eqnarray}
H^{\rm eff}_{\rm boson}&=& \frac{1}{2} \sum_{i,j} J^{\rm eff}(\br_i-\br_j)
\left[
\la B^\dagger_{ij} \ra B^{\vphantom\dagger}_{ij}
+ B^\dagger_{ij} \la B^{\vphantom\dagger}_{ij}\ra \right. \nonumber \\
&-& \left. \la A^\dagger_{ij} \ra A^{\vphantom\dagger}_{ij}
- A^\dagger_{ij} \la A^{\vphantom\dagger}_{ij}\ra \right]
\end{eqnarray}
(iii) Finally, we
take the constraint (\ref{constraint})
into account using a site-independent Lagrange multiplier $\lambda$,
thus working with the Hamiltonian $H^{\rm mf}_{\rm boson}
= H^{\rm eff}_{\rm boson} - \lambda \sum_i
\left( b^{\dagger}_{i\alpha} b^{\vphantom\dagger}_{i\alpha} - 2S \right)$.
The Hamiltonian $H^{\rm mf}_{\rm boson}$ is
readily diagonalized by a Bogoliubov rotation
\bea
b^{\vphantom\dagger}_{\bk,1} = u_\bk f^{\vphantom\dagger}_{\bk\upa}
- v_\bk f^{\dagger}_{-\bk\dna} \\
b^{\dagger}_{-\bk,2} = -v^*_\bk f^{\vphantom\dagger}_{\bk\upa}
+ u^*_\bk f^{\dagger}_{-\bk\dna}.
\eea
Defining
\bea
\beta_\bk &=& \sum_\br J^{\rm eff}(\br) B(\br) \cos(\bk\cdot\br)\\
\alpha_\bk&=& \sum_\br J^{\rm eff}(\br) A(\br) e^{-i\bk\cdot\br}\\
A(\bk) &=& \sum_\br A(\br) e^{i \bk\cdot\br}\\
B(\bk) &=& \sum_\br B(\br) e^{i \bk\cdot\br},
\eea
and choosing
\bea
\Omega_\bk&=&\sqrt{\vert\beta_\bk-\lambda\vert^2-\vert
\alpha_\bk\vert^2} \\
u_\bk&=&\cosh(\theta_\bk) e^{-i\gamma_\bk} \\
v_\bk&=&\sinh(\theta_\bk) e^{-i\gamma_\bk} \\
\cosh(2\theta_\bk)&=& \vert\beta_\bk-\lambda\vert/\Omega_\bk \\
\sinh(2\theta_\bk)&=& \vert\alpha_\bk\vert/\Omega_\bk.
\eea
we arrive at the diagonal form
\bea
\!\!\!\!H^{\rm mf}_{\rm\small boson} &\!\!=\!\!& \!\!\frac{1}{2}\!\!
\sum_{\bk,\mu} \Omega_{\bk}
f^\dagger_{\bk,\mu} f^{\vphantom\dagger}_{\bk,\mu} +
\lambda (S+1/2) N_{\rm site} \nn\\
\!\!\!\!&-&\frac{1}{2}\!\! \sum_\bk\! \beta(\bk)
+\frac{1}{2}\!\! \sum_\bk \big( A^*(\bk) \alpha_\bk
\!\!-\!\! B^*(\bk) \beta_\bk\big)
\label{diagbosonmft}
\eea
The $f$-particles appearing here are bosonic $S=1/2$ {\it spinons}.
The above parameters $u_\bk,v_\bk$ depend
on the mean field values of $\la B_{ij} \ra \equiv B(\br) $,
$\la A_{ij} \ra
\equiv A(\br)$,
and $\la \bS_i \cdot \bS_j\ra$ appearing in the quadratic Hamiltonian.
These mean field parameters
are evaluated in the ground state of (\ref{diagbosonmft}),
which results in a self-consistent theory.
For completeness, the self consistency conditions and the
constraint equation are given by
\bea
A(\bk)&=&\frac{\alpha(\bk)}{2\Omega_\bk} \\
B(\bk)&=&\frac{\beta_\bk-\lambda}{2\Omega_\bk} -\frac{1}{2} \\
2 S + 1 &=& \frac{1}{N_{site}} \sum_\bk 
\frac{\beta_\bk-\lambda}{\Omega_\bk}
\eea
It is well known \cite{sbreview}
that long-range magnetic order appears in this formulation if
the energy $\Omega_{\bk,\mu}$ vanishes at some
wavevector(s) $\bk=\{\bQ_i\}$ in the
thermodynamic limit, leading to condensation of the spinons at
these momenta. In the N\'eel state
on the square lattice, the spinons condense at wavevectors
$\bQ=\pm(\pi/2,\pi/2)$ in the thermodynamic limit \cite{gauge}
leading to magnetic
order at the wavevector connecting these points, namely $(\pi,\pi)$.
We report below the ground state properties of the
model with four-spin interactions, and present a comparison with
exact diagonalization results on small clusters.

\section{Ground state properties}
\subsection{Comparison with exact diagonalization results on small 
clusters}
Schwinger boson mean field theory
provides very accurate results for the ground state properties of
the nearest-neighbor Heisenberg antiferromagnet \cite{sbreview}. In order
to assess the accuracy of this approximation scheme in presence
of the four-spin interaction term it is useful to compare the mean-field
estimates of ground state properties with exact diagonalization numerics.
To this end, we have performed Lanczos exact diagonalizations of the
spin Hamiltonian (\ref{hspin}) on the $4\times 4$, 
$\sqrt{20}\times\sqrt{20}$,
$\sqrt{32}\times\sqrt{32}$ and $6\times 6$ clusters. In particular we have
focused on exchange interactions corresponding to
$U/t \to \infty$ (the Heisenberg limit), and $U/t=6.5$, for which
the mean field theory fits the experimental magnon dispersion of
La$_2$CuO$_4$ as discussed in the next section.
The latter choice of $U/t$ corresponds to $J_2 = J_3 = 0.0276 J_1 $
and $J_\Box = 0.55 J_1$ in (\ref{hspin}).
\begin{figure}
\begin{center}
  \centerline{\includegraphics[width=0.9\linewidth]{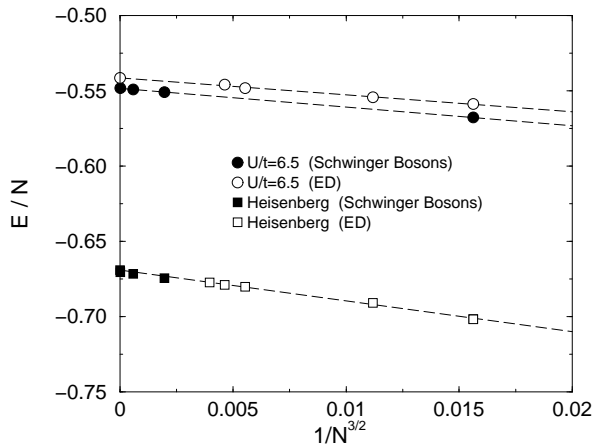}}
  \caption{Comparison of Schwinger boson mean field theory and
    exact diagonalization calculations of the
    ground state energy per site for the spin model in Eq.~(\ref{hspin})
    for the Heisenberg limit ($U/t \!\to\!\infty$)
    and for $U/t=6.5$.}
  \label{fig:egs-comparison}
\end{center}
\end{figure}
\begin{figure}
\begin{center}
\centerline{\includegraphics[width=0.9\linewidth]{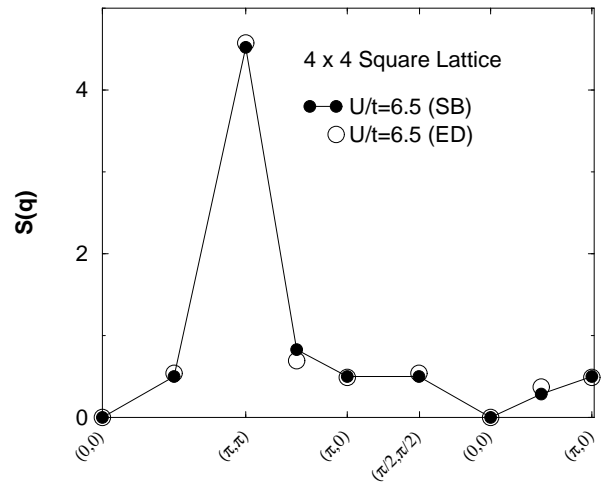}}
\caption{Comparison of Schwinger boson mean field theory 
and exact
diagonalization calculations of the
static (equal-time) spin structure factor $S(\mathbf{q})$ on a 
$4\times 4$ lattice. These results are for the bare spin-spin
correlation function, where the transformation of the spin
operator in going from the Hubbard model to the spin model 
have not been included --- including them modifies the mean-field
and the exact diagonalization results in the same manner, as discussed
in the text, and does not affect our comparison.}
\label{fig:lanczos-sq}
\end{center}
\end{figure}

We have calculated the ground state energy and the static spin structure
factor $S(\bq)$
\be
S(\bq)= \sum_{j} \la \bS(\br_j)\cdot\bS(\br_0) \ra
e^{-i \bq \cdot (\br_j-\br_0)}
\label{spinsq}
\ee
The results presented in Figs.~\ref{fig:egs-comparison} and
\ref{fig:lanczos-sq}
clearly show that the Schwinger boson approach is still rather accurate 
even
for large four-spin interactions which are generated for a moderate 
interaction strength $U/t=6.5$.
We next turn to the staggered magnetization and spin stiffness
in the thermodynamic limit for this parameter value, taking into
account the modification of the spin operator by local charge
fluctuations.

\subsection{Staggered magnetization and spin stiffness in the 
thermodynamic
limit}
As mentioned earlier, the local charge fluctuations in the Hubbard model
affect the spin correlations. We can take this into account by
noticing that the spin operator of the Hubbard model is modified
by the same unitary transformation which enables us to rotate from
the Hubbard model Hamiltonian to the Heisenberg model with four-spin
interactions of Eq.~(\ref{hspin}) which has no charge fluctuations. 
This transformation was discussed earlier in the
context of various physical correlations in the doped cuprate 
superconductors \cite{vmclong}, 
and its importance for spin physics in the undoped
Mott insulator was pointed out by Delannoy {\it et al}.\cite{tremblay}

For instance, making the unitary transformation on the spin operator 
$S^+(\br)$, we find
\bea
\tilde{S}^+(\br) &=& S^+(\br) (1- 4\frac{t^2}{U^2}) + \frac{t^2}{U^2} 
\sum_{\alpha=\pm\hx, \pm\hy} S^+(\br+\alpha) \nonumber \\
&-& \frac{t}{U} \sum_{\alpha=\pm\hx, \pm\hy} \left( 
c^{\dagger}_{\br+\alpha,\upa} c^{\vphantom\dagger}_{\br,\dna} - 
c^{\dagger}_{\br,\upa} c^{\vphantom\dagger}_{\br+\alpha,\dna} \right)
\label{spinopmodify}
\eea
While the first two terms lead to spin flips within the lower Hubbard
band, the last term excites an electron into the upper Hubbard
band. This final term plays a role at high energies, of ${\cal O}(U)$,
and short distances \cite{tremblaycomment}. 
We will therefore not consider it below, but
restrict attention to energies well below the Mott gap, where the
first two terms alone play a role and our spin-only description is
valid. With this caveat, we find at ${\cal O}(t/U)^2$ we find that
the transformed ``low energy'' equal-time spin-spin
correlation function is given by
\bea
\tilde{\cS}_{\rm low} (\br-\br') 
&= &
\la \bS(\br)\cdot \bS(\br')\ra (1-8 \frac{t^2}{U^2}) \nonumber\\
&+& \frac{t^2}{U^2} \sum_{\alpha=\pm\hx,\pm\hy} \la \bS(\br+\alpha)\cdot
\bS(\br')\ra \nonumber\\
&&\qquad \qquad  + \la \bS(\br) \cdot \bS(\br'+\alpha)\ra
\label{spinmodify}
\eea

Within Schwinger boson mean field theory, we obtain the antiferromagnetic
order parameter $m_s$
from finite-size scaling of the static structure factor, defined
via
\be
\tilde{\cS}_{\rm low} (\bq)= \sum_{j} \tilde{\cS}_{\rm low} (\br_j-\br_0)
e^{-i \bq \cdot (\br_j-\br_0)},
\label{spinmodsq}
\ee
as
\be
m^2_s =\lim_{N_{\rm site}\to\infty} \frac{\tilde{\cS}_{\rm low}
(\pi,\pi)}{N_{\rm site}}.
\ee
We find that the staggered magnetization for $U/t=6.5$ extrapolates
to $m_s \approx
0.25$, smaller than that for the nearest-neighbor Heisenberg
model ($m_s^{nn} \approx 0.303$). This decrease of the staggered 
magnetization
is due to local charge fluctuations in the insulator, and arises
from ${\cal O}(t/U)^2$ terms in the modified spin operator in
Eq.~(\ref{spinopmodify}) as pointed out in Ref.~\onlinecite{tremblay}.

For a pure spin model of the copper sites in La$_2$CuO$_4$, we 
expect the sum-rule on the equal-time structure factor $\sum_\bq 
S(\bq) = S(S+1)$ with $S=1/2$. However, with the above modification
to the spin correlations, this sum-rule is corrected by terms of
order $(t/U)^2$. Specifically, imagine integrating the dynamical
structure factor over energies relevant to neutron scattering
experiments, i.e. much smaller than the Mott gap, and then integrating
this ``low-energy'' equal-time structure factor $\tilde{\cS}_{\rm low}
(\bq)$ over all $\bq$. In this case, we find the modified relation
$\sum_\bq \tilde{\cS}_{\rm low} (\bq) = 
S_{\rm eff} (S_{\rm eff}+1)$ with $S_{\rm eff} \simeq
0.39$. It would be interesting to look for this apparent sum-rule
deficit in neutron scattering studies.

Following Stringari \cite{stringari}, we extract the spin stiffness
from the small momentum behavior
of the structure factor. Specifically, approaching the magnetically
ordered state in a rotationally invariant formulation, the static
structure factor for small $q$ is related to the spin stiffness through
\be
\rho_s = \frac{c \tilde{\cS}(\bq \to 0)}{q}
\label{stringari}
\ee
where $c$ is the spin wave velocity of the linearly dispersing
magnon at momenta near $(0,0)$ and $(\pi,\pi)$. The spin-wave
velocity $c$ can be obtained from the magnon dispersion as discussed
in connection with the excitation spectrum in the next section.
This, together with the calculated structure factor in the limit,
$\bq \to 0$, yields the spin stiffness through Eq.~(\ref{stringari}).
As discussed in detail in the next section, a fit to
the overall magnon dispersion in La$_2$CuO$_4$
yields a spin wave velocity $c \approx 800 meV$-$\AA$. Assuming a
lattice spacing $3.85\AA$
for the CuO$_2$ plane, this yields $\rho_s \approx 24 meV$,
which is consistent with experimental findings.
\cite{neutrons}
Note that the spin
stiffness can also be obtained by studying the change in ground
state energy for a slowly varying spin twist.
\section{Excitation spectrum}
To probe the spectrum of $S=1$ excitations at a given momentum $\bq$,
which is of interest for neutron scattering experiments, one can act
with the operator $S^+(\bq) = \sum_\bk b^\dagger_{\bk,1}
b^{\vphantom\dagger}_{\bk-\bq,2}$ onto
the ground state \cite{stringari,auerbachbook}. Making a Bogoliubov 
rotation to
spinon variables, we can rewrite this as
\bea
S^+(\bq) = \sum_\bk \left(u_\bk f^{\vphantom\dagger}_{\bk\upa} -
v_\bk f^\dagger_{-\bk\dna} \right) \nonumber \\
\times
\left(u_{\bk-\bq} f^{\vphantom\dagger}_{\bk-\bq,\upa} -
v_{\bk-\bq} f^\dagger_{-\bk+\bq,\dna} \right).
\label{spinexc}
\eea

Including corrections to the spin operator via the unitary
transformation, discussed earlier
for the static structure factor, leads to two modifications
in the dynamical response. The
first is a simple modification of the spin operator
by a multiplicative
``form factor'' $S^+(\bq) \to G(\bq) S^+(\bq)$, with
\be
G(\bq) = 1 - \frac{2 t^2}{U^2} \left( 2 - \cos q_x - \cos q_y \right)
\ee
This will not affect the magnon energies but will correct the dynamical
response by an overall multiplicative prefactor $G^2(\bq)$.

The second effect of this unitary tranformation as discussed
is to generate a new term
which causes a transition from the
lower to the upper Hubbard band. This term plays a role only if we
examine spin dynamics at very high energies, and is of no relevance
at energies probed in the neutron scattering experiments to which
we restrict attention here.
We see that at a given momentum $\bq$, if one of the two spinons
combining to give the spin operator in Eq.~(\ref{spinexc})
is condensed (which would happen
for spinon momenta $\pm(\pi/2,\pi/2)$ in the N\'eel state),
that part
behaves as a single particle excitation with a well defined dispersion
 - this is the magnon. For
general $\bk$, both spinons are uncondensed and this remainder of
the sum contributes to the scattering continuum. Quite generally,
both parts play a role when we evaluate the spectral function for
the $S^+(\bq)$ operator.

\subsection{Magnon dispersion}
From the above discussion, it is clear that the magnon energy
at momentum $\bq$ within the mean field theory of the N\'eel state
is just given by $\Omega_{\bQ-\bq}/2$ where $\bQ=\pm(\pi/2,\pi/2)$.
For our choice of mean field decoupling, and for spin
$S \to \infty$, this is exactly half the value given by spin-wave
theory \cite{spinwave}. It is known \cite{auerbachbook}
that Schwinger boson mean field theory for the Heisenberg model
has this shortcoming, which can be fixed by working within
a large-$N$ generalization of the Schwinger boson theory and
including $1/N$ corrections to the mean field ($N=\infty$)
result. Since we reduce our original Hamiltonian in Eq.~(2)
to an effective
Heisenberg-type model before we use the Schwinger boson representation,
we expect the same fluctuation corrections to appear in our case as well,
however an explicit calculation is beyond the scope of this
paper. Here and below we will work with $\Omega_\bk$ as the magnon energy.
Note that the ${\cal O}(t/U)^2$ corrections to the spin operator do not
affect the magnon dispersion.
\begin{figure}
\begin{center}
\centerline{\includegraphics[angle=-90,width=0.9\linewidth]{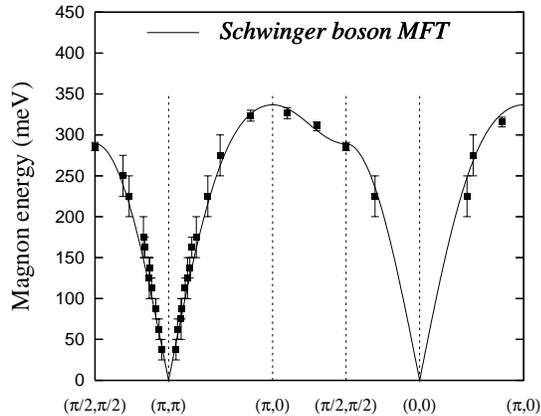}}
\caption{Fit of the magnon dispersion obtained from Schwinger boson mean
field theory to the experimental data at $T=10K$
in Ref.~\onlinecite{coldea} along
the standard contour in the Brillouin zone. The
Hubbard model parameters used for this fit are $t =360 meV$ and $U=2.34 
eV$,
with the effective spin Hamiltonian couplings given by Eq.~(\ref{Jexpr}).}
\label{fig:magnondispersion}
\end{center}
\end{figure}
Fig.~\ref{fig:magnondispersion} shows the magnon energy
$\Omega_\bk$ obtained within the mean field theory for $U/t=6.5$
and $t=360meV$ compared with the experimental data from Coldea
{\it et al}. This choice of parameters gives a good fit to the
dispersion over the entire contour along the Brillouin zone,
including both the linearly dispersing spin-wave regime as well as
the high energy magnon dispersion along the zone boundary. While
the fit along the zone boundary is easy to see in the figure, its
accuracy at low energies may be seen from the spin wave velocity
in the mean field theory $c\approx 800 meV$-$\AA$ being in good
agreement with experiment. We also point out that at much larger
values of $U/t$ ($\gtrsim 7$), we are unable to reproduce properly
the dip in the magnon dispersion at $(\pi/2,\pi/2)$ while
traversing the zone boundary. While further neighbor hoppings in
the Hubbard model (e.g. third-nearest neighbor hopping $t_3$)
could produce a similar dip near $(\pi/2,\pi/2)$ by generating an
antiferromagnetic exchange of order $t^2_3/U$, electronic
structure calculations for La$_2$CuO$_4$ indicate that such
hopping matrix elements are too small to be relevant.

\subsection{Scattering continuum}
\begin{figure}
\begin{center}
  \centerline{\includegraphics[width=0.9\linewidth]{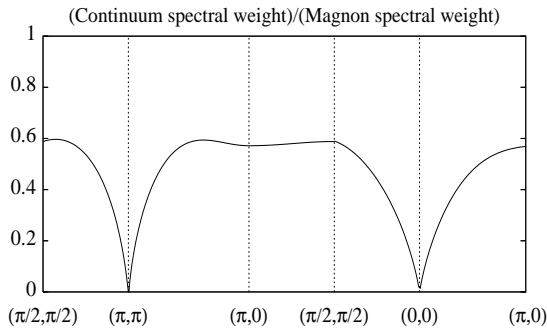}}
  \caption{Ratio of continuum and magnon spectral weights
    along a contour in the Brillouin zone. The magnon
    exhausts the spectral weight for $\bq \to (0,0)$ and $\bq \to
    (\pi,\pi)$ consistent with general arguments \cite{stringari}.
    Along the zone boundary, the continuum accounts for nearly
    $40\%$ of the total spectral weight.}
  \label{fig:spwt}
\end{center}
\end{figure}

\begin{figure}
\begin{center}
\hspace*{0mm}
\centerline{\fig{3.0in}{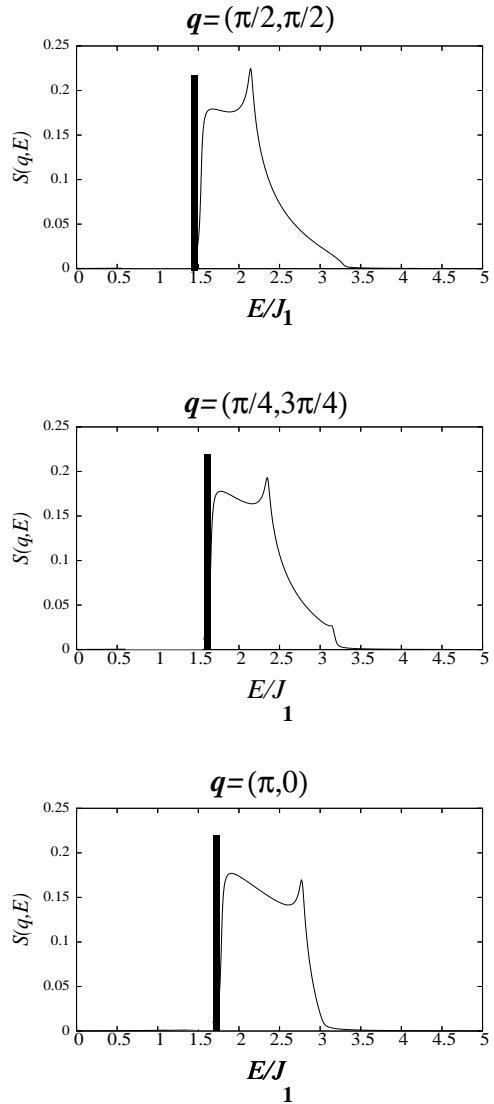}}
\caption{The dynamical structure factor along the zone boundary
at $(\pi/2,\pi/2)$ (top), $(\pi/4,3\pi/4)$ (middle) and
$(\pi,0)$ (bottom), showing the sharp magnon (as a bold vertical
line not drawn to scale) and the broad continuum including the sharp
secondary features most clearly visible at $(\pi/4,3\pi/4)$.
%There is no spectral weight below the magnon peak - its presence in
%all three plots is only an artifact of the Lorentzian
%broadening of the continuum spectra used in the numerics.
}
\label{fig:continuum}
\end{center}
\end{figure}

Having fixed the parameter values $U,t$ from fitting the magnon
dispersion to the data, we next turn to the scattering continuum.
Specifically, we are interested in asking how much weight is
present in the continuum relative to the magnon, and if there are
any features in the continuum which may be experimentally observable.
Recall that in the mean field theory this part of the spectrum
is just the two-spinon excitation continuum.
In order to obtain the weight in the continuum versus the magnon,
we consider the energy integrated response. This is
just the structure factor
\bea
S(\bq)&=& \sum_\bk \left[ \left(\frac{\vert\beta_{\bk-\bq}-\lambda\vert}
{\Omega_{\bk-\bq}} -1 \right)
\left(\frac{\vert\beta_\bk-\lambda\vert}{\Omega_{\bk}}
+ 1 \right) \nonumber \right. \\
&-& \left. \frac{\alpha^*_\bk \alpha_{\bk-\bq}}{\Omega_\bk
\Omega_{\bk-\bq}} \right]
\eea
To obtain the magnon weight part, we need to keep only those
contributions where $\Omega_\bk$ or $\Omega_{\bk-\bq}$ would
vanish in the thermodynamic limit (or are minimum on a finite
but large lattice). The continuum contribution is the remaining
part of the sum. Fig.~\ref{fig:spwt} shows the ratio of these
weights along the same contour in the Brillouin zone over which
the magnon dispersion is displayed. This ratio is clearly insensitive
to the ``form factor'' $G(\bq)$, however
the important thing to note from the plot is that the continuum
weight is most significant along the zone boundary and accounts
for nearly $40\%$ of the total spectral weight. This is the
most promising region to study the continuum in neutron scattering
experiments. The continuum weight is also considerable along the line
from $(\pi,0)$ to $(\pi,\pi/2)$. This is qualitatively similar
to results inferred for the Heisenberg model from QMC calculations
\cite{sandvik}. 
%although we cannot make any quantitative comparisons
%between our results.
\begin{figure}
\begin{center}
  \centerline{\includegraphics[width=\linewidth]{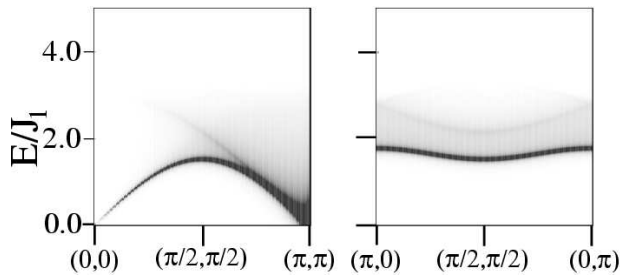}}
  \caption{Grayscale plot of the dynamical structure factor $S(\bq,E)$
    of the spin model (\ref{hspin}) with optimal parameters as in Fig.~1,
    with high intensity in black. The gray areas indicate regions of
    continuum scattering. The intensity has been scaled to clearly show
    both the magnon and secondary peak features. \underline {Left panel}:
    Magnon peak and the single secondary peak along $(0,0) \to (\pi,\pi)$.
    The magnon and secondary peak intensities vanish as $\bq \to (0,0)$.
    \underline{Right panel}: Magnon peak and two secondary peaks along the
    zone boundary
    $(\pi,0) \to (0,\pi)$. The two secondary peaks merge at $(0,\pi)$
    and $(\pi,0)$. The dip
    in the magnon dispersion at $(\pi/2,\pi/2)$ is clearly visible.}
  \label{fig:SQE}
\end{center}
\end{figure}
In order to display possible interesting features in the
continuum scattering we have plotted the spectrum of
excitations at a few $\bq$ points along the zone boundary
in Fig.~\ref{fig:continuum},
as well a grayscale plot of the spectral function in
Fig.~\ref{fig:SQE}
along $(0,0) \to (\pi,\pi)$ and along $(\pi,0)\to(0,\pi)$.
We see that in addition to the magnon, there are singular
secondary features in the spectrum. These secondary peaks
can be shown to arise from special points $\{\bK\}$ in the
Brillouin zone in the vicinity of which the sum $\Omega_\bk+
\Omega_{\bk-\bq}$ varies slowly (dispersing as $\sim \vert
\bk -\bK\vert^4)$ giving rise to a log singular density of states
for the two-spinon continuum.
These secondary peaks and continuum scattering also arise within the
mean field theory for the nearest neighbor Heisenberg model,
with minor quantitative changes due to differences in
the spinon dispersion. The effect
of fluctuations beyond the mean field result on the magnon
dispersion and the scattering continuum is beyond the scope
of this paper, and is currently being investigated.
\section{Concluding Remarks}
We have presented a Schwinger boson mean-field description of the
of the effect of four-spin interactions on spin dynamics in Mott
insulators. Such multispin interactions
and their effects are potentially important in any correlated
insulator with local charge fluctuations. For
La$_2$CuO$_4$, the experimental magnon dispersion is well reproduced by
our approach with Hubbard model parameters $U \approx 2.34 eV$,
and $t \approx 360 meV$. These values are consistent with
the experimental estimates \cite{optics}, and
electronic structure calculations \cite{estruct}.
This leads to a sizeable value of the four-spin exchange term
$J_\Box \approx 0.55 J_1$ in the effective spin Hamiltonian.
Good agreement with experiments is also obtained for
spin wave velocity and spin stiffness,
thus leading to a consistent description of the magnetic behavior of
this Mott insulator. The staggered moment is substantially
reduced from that for the Heisenberg model, it would be worth
revisiting experiments to test for this reduced ordered moment.
We have also discussed differences in sum rules between Hubbard
and Heisenberg models, and refer the reader
to more recent work \cite{lorenzana} for additional insights,
alternative approaches, and the current experimental status.
Beyond the single magnon excitation,
we have shown that there is considerable spectral weight
in the continuum along the zone boundary,
and that the dynamical structure factor exhibits secondary
peaks which we understand as arising from the density of states of
two-spinon excitations. These could be explored in experiments
with high intensity neutron sources in the near future.
%%\vspace{1cm}
%%\ni{\bf Acknowledgments}
\acknowledgments
We thank Radu Coldea, Gregoire Misguich and Prof. Andre-Marie 
Tremblay for very useful discussions and correspondence, and Anton
Burkov for an early stimulating conversation. AP acknowledges
support through grant DOE LDRD DEAC03-76SF00098. The numerical
calculations have been performed on the ``Mizar'' SMP machine
at EPFL.


\begin{thebibliography}{999}
\bibitem{gauge}
This is for our particular mean field decoupling. The spinon dispersion
is not a gauge invariant quantity, although the number of gapless
wavevectors and the momentum difference between them are gauge invariant.
\bibitem{tremblaycomment}
This term was inadvertently omitted in the earlier version of this
paper, we are indebted to Prof. Andre-Marie Tremblay for
pointing this out to us.
\end{thebibliography}
\end{document}